# Wide band gap tunability in complex transition metal oxides by site-specific substitution

Woo Seok Choi, Matthew F. Chisholm, David J. Singh, Taekjib Choi, Gerald E. Jellison, Jr. and Ho Nyung Lee*

Materials Science and Technology Division, Oak Ridge National Laboratory, Oak Ridge, TN 37831, USA.

*e-mail: hnlee@ornl.gov.

**Fabricating complex transition metal oxides with a tuneable band gap without compromising their intriguing physical properties is a longstanding challenge. Here we examine the layered ferroelectric bismuth titanate and demonstrate that, by site-specific substitution with the Mott insulator lanthanum cobaltite, its band gap can be narrowed as much as one electron volt, while remaining strongly ferroelectric. We find that when a specific site in the host material is preferentially substituted, a split-off state responsible for the band gap reduction is created just below the conduction band of bismuth titanate. This provides a route for controlling the band gap in complex oxides for use in emerging oxide opto-electronic and energy applications.**

Band gap tuning is at the core of current materials research and opto-electronic device applications. Through successful tuning of the band gap in semiconductors, band gap tailored heterostructures including 2D electron gas and tunnelling structures were realised. Such advancements offered a greater understanding of physics regarding quantum electrodynamics and stimulated emergence of many related devices[1]. Moreover, the ability to tune the band gap is becoming increasingly important for developing highly efficient solar cells and transparent conducting oxides. For conventional III-V and II-VI semiconductors, simple band gap tuning has been extremely successful leading to realisation of the structures mentioned above. For example, GaAs has a band gap of 1.42 eV, which can be continuously tuned down to 0.35 eV or up to 2.12 eV by substituting In or Al for Ga, respectively. Such simple alloying results in a band gap spectrum of larger than one electron volt. On the other hand, recent breakthroughs in complex oxides have provided an opportunity to incorporate our understanding of semiconductors into the exotic physics of transition metal oxides (TMOs)[2,3]. For example, the observation of quantum transport behaviours in several complex oxides manifest a substantial improvement of oxides in terms of their quality and leading to properties that were thought to be unique to semiconductors[4,5]. However, substantial and controllable band gap tuning has yet to be achieved in TMOs, despite intense effort. In order to tune the band gap of TMOs, one might consider modifying or substituting the transition metal (TM) with another element, because the rigid nature of the band gap originates mostly from the strongly localised character of the $d$-electrons. However, what has been observed is that the fascinating physical properties of TMOs arising from the $d$-electrons disappear with the modified band gap. Such difficulties have hampered the recent searches for more efficient transparent conducting oxides and low band gap photovoltaic oxides[6-11]. Ferroelectric oxides, in particular, are attracting renewed attention owing to their inherent built-in potential arising from the spontaneous polarization which could be used to efficiently separate photon-induced electron-hole pairs[7-11].

In order to identify a useful route for effectively controlling the band gap in complex oxides, we focused on the ferroelectric $Bi_4Ti_3O_{12}$ (BiT), which has an optical band gap energy reported between 3.1 and 3.6 eV[12-14]. Unlike typical ferroelectric perovskites such as $BaTiO_3$, $Pb(Zr,Ti)O_3$, or $BiFeO_3$, BiT has an Aurivillius phase with a highly anisotropic monoclinic (pseudo-orthorhombic, $a$ = 5.450 Å, $b$ = 5.406 Å, and $c$ = 32.832 Å)[15] layered structure with two alternating atomic arrangements; one fluorite-like $(Bi_2O_2)^{2-}$ layer and two perovskite $(Bi_2Ti_3O_{10})^{2+}$ layer blocks connected with an extra oxygen plane, as schematically shown in Fig. 1a. Note that the $c$-axis lattice constant of BiT is about eight times longer than that of usual simple perovskites. The ferroelectric polarization in BiT mainly points along the $a$-axis direction[16] and is known to originate mostly from the perovskite sub-layers, *i.e.*, the shift of cations relative to O ions within the perovskite blocks[17,18]. One of the most attractive properties of this bismuth-layered material is its excellent sustainability of ferroelectricity even after intrusion of some oxygen vacancies[17]. (Note that simple perovskite-typed ferroelectrics are usually less tolerant to oxygen vacancies.) Thus, the unique layered structure of BiT provides an opportunity to examine site-specific substitutional alloying for altering the band gap without losing the ferroelectricity. The alloying material was selected among simple perovskite TMOs with low band gap and high absorption ($\alpha$) above the band gap. La$TM$O$_3$ was our natural choice for the perovskite, since La doping is well known to stabilise the ferroelectricity in BiT[19,20]. Among La$TM$O$_3$, $LaCoO_3$ (LCO) is chosen as the alloying material. LCO is a Mott insulator with a small band gap of 0.1 eV and also a high $\alpha$ originating from the Co 3$d$-electronic states[21]. It also has a rhombohedral structure ($a$ = 5.378 Å) with an in-plane lattice mismatch with BiT less than 1%, readily allowing coherent incorporation into BiT without destroying the overall layered structure.

In this article, we present a combined experimental and theoretical study on LCO-alloyed BiT (BiT-LC) to elucidate the possiblity of band gap tuning in complex transition metal oxides. Through site-specific substitutional alloying, we successfully reduced the band gap as much as ~1 eV, without deterioration of the ferroelectricity of BiT. Moreover, we observed enhanced photoresponses from the alloyed BiT thin films, clearly confirming the band gap reduction.

## Results

**Thin film samples.** We used pulsed laser epitaxy (PLE, also known as pulsed laser deposition) to fabricate high-quality BiT-LC epitaxial thin films on SrTiO$_3$ (STO) substrates. Details on the film growth and other experimental setups are given in the Methods section. The fabricated BiT-LC samples were intended to have either one or two pseudo-cubic unit cell layers of LCO incorporated into one unit cell layer of BiT, which hereafter will be nominally called as 1Bi$_4$Ti$_3$O$_{12}$-1LaCoO$_3$ (1B1L) or 1Bi$_4$Ti$_3$O$_{12}$-2LaCoO$_3$ (1B2L), respectively. (However, in reality, much less Co was incorporated in the films relative to the La substitution as discussed later.) For higher concentrations of LCO alloying, coherent incorporation was not obtained, resulting in a severe degradation of the unique layered structure of BiT.

**Site-specific substitution: Atomic structure and chemical analysis.** The crystal structure of the BiT-LC thin films were characterised using X-ray diffraction (XRD) and Z-contrast scanning transmission electron microscopy (STEM), showing clear evidence of substitutional alloying. Figures 1b and 1c show XRD $\theta$-$2\theta$ and rocking curve scans, respectively, from BiT-LC thin films. The well-defined thickness fringes observed for most of the peaks in $\theta$-$2\theta$ patterns together with very narrow full width at half maxima (FWHM, $\Delta w$ ~0.025°) in rocking curve scans indicate an atomically smooth surface and interface as well as excellent crystallinity of our thin films. (Note that the FWHM in XRD rocking curves in our films reported here is drastically reduced at least by a factor of three as compared to the values from previous epitaxial films[20]) Overall peak positions of the main peaks remain unchanged for all the samples up to 1B2L, indicating that the overall Aurivillius structure of BiT is well preserved even after the substitution. This suggests that La and Co were successfully incorporated in BiT. Z-contrast STEM investigations confirm this substitutional alloying, where the characteristic stacking of BiT layers is clearly preserved in BiT-LC as shown in Fig. 1d.

A more crucial observation obtained from STEM is the signature of site-specific alloying. Figure 1e shows a magnified annular dark field (ADF) image along with the elemental maps of Ti and La obtained by electron energy-loss spectroscopy (EELS) in a 1B2L sample. The ADF image mostly shows bright features at the Bi and La sites with reduced intensity from the Ti sites. The alternating $Bi_2O_2$ and $Bi_2Ti_3O_{10}$ sub-layers are clearly seen. The Ti map obtained using the Ti-$L$ edge shows that Ti is localised within the $Bi_2Ti_3O_{10}$ sub-layer only. The La map indicates that asymmetric La substitution for Bi with preferential substitution into the upper layer of Bi within the $Bi_2O_2$ sub-layer. While considerably less Co incorporation in BiT is detected by STEM-EELS, Co is seen to substitute for Ti. The small amount of Co substitution makes it difficult to determine if it, like La, is asymmetrically incorporated in BiT. However, as discussed later, it is suggested that the La and Co substitution in BiT mostly takes place near the $Bi_2O_2$ sub-layer region, forming a site-specifically alloyed, self-ordered superstructure. The reduced incorporation of Co relative to La in our thin films seems to originate from the energetically unfavourable electronic state of Co and the different adsorption rate of the elements during the deposition. Since Co has two primary valence state in oxides ($Co^{3+}$ and $Co^{2+}$), where $Co^{2+}$ is generally more stable, Co substitution in either case for $Ti^{4+}$ is more difficult as it requires a compensation of the charge difference, while $La^{3+}$ could readily substitute for volatile $Bi^{3+}$. Nevertheless, the impact of Co on the band gap change is amazingly large as discussed later. Moreover, it has been reported that O vacancies readily form in the Aurivillius structure and prefer sites near the $Bi_2O_2$ layer, without compromising the ferroelectricity[17,22,23]. Therefore, further considering the difference in ionic size as well as electronegativity, we presume that Co preferentially substitutes for Ti next to the $Bi_2O_2$ layer accompanying O vacancies to compensate valence difference.

Site-specific substitution of LCO is further supported by the XRD $\theta$-$2\theta$ patterns in Fig. 1b. In the XRD $\theta$-$2\theta$ patterns, it should be noted that the peak intensity of higher order peaks (006 and 008) stays almost the same with increasing La and Co substitution as the ones from pure BiT, but that of lower order peaks (002 and 004) decreases as shown in Fig. 1b. This indicates that the short period perovskite structure is conserved upon alloying, but the long period $Bi_2O_2$ sub-layer seems to be disturbed. Another important observation is the separation of the 004 peak in all of our samples, as indicated with the arrows in Fig. 1b. While the origin of the peak separation is not yet clear, it suggests that there is a change in the $c/4$ lattice of BiT. This might originate from the fact that the distance between upper and lower Bi to O within the $Bi_2O_2$ layer is slightly different; the difference between the quarter of the $c$-axis lattice constants (c'/4 and c"/4) is 0.03 nm for the pure BiT film, as schematically shown in Fig. 1a. The separation of the 004 peak increases when LCO alloying is introduced, where the c/4 lattice constant difference further increases to ~0.06 nm for both 1B1L and 1B2L. The increase of modulation by alloying LCO accounts for the observation in STEM where La preferentially substitutes in the upper Bi layer within the $Bi_2O_2$ sub-layer in BiT.

**Persistent ferroelectricity.** In order to check the effect of the substitution on the ferroelectric properties of BiT-LC films, we have grown (104)-oriented films on STO (111) covered with $SrRuO_3$ bottom electrodes. (It was rather difficult to characterise the ferroelectric properties from a $c$-axis oriented film since its primary polar axis is lying along the in-plane direction[24].) The polarization of a (104)-oriented film is tilted by 54.7° from the normal of the film plane. We measured the polarization ($P$) as a function of electric field ($E$) as shown in Fig. 1f. Surprisingly, the overall shape of the $P(E)$ loop from BiT-LC is almost identical to that from the pure BiT, indicating its persistent ferroelectricity even after the substitution. The remnant polarization is ~15 $\mu C/cm^2$ for both BiT and BiT-LC, comparable to the previously reported values of 15.9 $\mu C/cm^2$ from a (104)-oriented La doped BiT film[24] (Capacitance and dielectric loss measurements are shown in Supplementary Fig. S1 and further confirm the excellent ferroelectric properties of our thin films). This is an important observation since alloying a simple perovskite-type ferroelectric with heterovalent TMs typically introduces extra charges that induce a (semi-)metallic state and severely degrade the ferroelectricity. Therefore, the persistent ferroelectricity in BiT-LC further supports the model of preferential substitution of $Bi_2O_2$ sub-layers based on STEM and XRD results described above: The $Bi_2Ti_3O_{10}$ sub-layers remain mostly intact to reveal stable ferroelectricity, even after alloying BiT with La and Co.

**Optical band gap control.** The optical conductivity as a function of photon energy ($\sigma_1(\omega)$) of our BiT-LC films was measured using spectroscopic ellipsometry as shown in Fig. 2a. While the largely undisturbed $Bi_2Ti_3O_{10}$ perovskite layers preserve the ferroelectricity in BiT-LC, the changes around the $Bi_2O_2$ sub-layer modify the optical properties substantially. Charge transfer excitations between O 2$p$ and Ti 3$d$ states are known to mainly contribute to the above band gap absorption of pure BiT[13]. Strong optical transitions start at ~3.55 eV for our BiT film, which we define as the band gap here in order to compare different thin film samples on an equal footing (see Ref. 13 for a detailed discussion of the spectra). For 1B1L and 1B2L, on the other hand, the band gap decreased drastically to 3.30 and 2.65 eV, respectively. This could be attributed to LCO that has a larger $\alpha$, compared to that of typical semiconductors as shown in Fig. 2b. Figures 2c and 2d show photographs of BiT and 1B2L on STO, respectively. While BiT is transparent, 1B2L shows a darker colour due to the enhanced absorption of light in the visible wavelength range. The dark colour is not related to unwanted oxygen vacancies typically found in reduced perovskite oxides.[25] In fact, all samples were highly insulating. Moreover, as explained in Supplementary Discussion, our optical spectroscopic analyses preclude the existence of unwanted Co-based phases, and their contribution to the band gap reduction.

The amount of band gap change is exceptionally large as compared to the reported values from other doping studies on ferroelectrics. For BiT, the largest band gap change reported so far was less than 0.2 eV, referenced to films with a reported gap of 3.64 eV[14]. Other ferroelectrics have similar rigid band gaps. For example, Pb(Zr,Ti)O$_3$ shows an experimental band gap spectrum less than 0.2 eV by doping.[26,27] Thus, we have accomplished wide band gap tunability (~1 eV), which is about five times greater than what has been reported so far from oxide ferroelectrics. Note that a recent density functional theory work has predicted that substituting the Ti in PbTiO$_3$ with another metallic element may enhance the polarization while reducing the band gap[10]. Results from an experimental examination with the exact materials' system, however, have not been reported. Thus, the significantly reduced band gap experimentally realised in our 1B2L films without reduction of polarization opens up the possibility of discovering new low band gap ferroelectric materials.

It should be noted that the key ingredient to modify the electronic structure of BiT-LC is Co. Although it is present at a smaller concentration than La, the substitution is surprisingly effective in reducing the band gap. Indeed, doping with only La in BiT is ineffective and known to rather slightly increase the band gap of BiT[28]. This has also been confirmed by testing the substitution with other elements (e.g., Ti and Al using LaTiO$_3$ and LaAlO$_3$, respectively) instead of Co. Alloying with these elements did not change the band gap of BiT at all (Supplementary Fig. S2). Therefore, it is noteworthy to point out that Co plays a critical role in manipulating the band gap of BiT.

**Electronic structure: Theory.** To understand the evolution of the electronic structure and the effectiveness of Co in reducing the band gap of BiT-LC, we performed density functional theory calculations. Following the observations from the experiment, we substituted Ti near the Bi$_2$O$_2$ sub-layer with Co in our calculation. We focused on Co$^{2+}$ in the outer perovskite block in Bi$_2$Ti$_4$O$_{10}$ layer combined with an O vacancy, which we took as the apical O on the outside of the perovskite block consistent with the crystal chemistry of $Bi_4Ti_3O_{12}$ (See Supplementary Methods for details). Figure 3b shows the calculation result from Co:BiT, in comparison with that for pure BiT in Fig. 3a. As usual, band gaps are underestimated with standard density functional calculations. The calculated band gap of 2.34 eV in pure BiT was reduced to 1.64 eV in the Co:BiT supercell, which is comparable to the 30% decrease observed by the ellipsometry. As shown in Fig. 3b, the band gap reduction in Co:BiT is mainly due to the split-off density of states in the conduction band (indicated as red arrow), meaning that there will be charge transfer excitations below the pure BiT gap. This is consistent with the additional absorption peak in ellipsometry data shown as red arrow in Fig. 2. The split-off states are mainly of Bi $6p$ character associated with the Bi atoms nearest the O vacancy. Therefore, the decreased band gap in BiT-LC can be attributed to the divalent Co incorporation into the BiT accompanied by O vacancies that pull Bi state down from the conduction band edge. This theoretical result shows that it is possible to obtain the electronic structure in accord with our experimental observations based on Co substitution combined with O vacancies. Such vacancies are to be expected in a Co-substituted material because of the chemistry of Co and, in particular, the fact that the most common oxidation states of Co are +2 and +3.

**Photoelectric response.** The photocurrent depends on a variety of factors including $\alpha$, band gap, carrier mobility, and light intensity. To see the effect of reduced band gap, we measured photocurrent as a function of $\omega$ by using a monochromatic light source in the in-plane geometry. As shown in Fig. 4a for 1B2L, the current density as a function of applied electric field ($J(E)$) for $\omega$ smaller than 2.76 eV shows similar curves as dark $J(E)$ indicating there is no absorption of light. Increased photocurrent starts to appear for $\omega$ larger than 2.82 eV. Figure 4b summarises the wavelength dependent photoconductance. $J$ for 1B2L starts to increase at 2.82 eV, whereas $J$ for pure BiT does not deviate from dark $J$ until $\omega$ reaches 3.18 eV, clearly confirming the enhanced $\alpha$ and photoelectric response due to the reduced band gap in BiT-LC. Note that the onset values approximately correspond to the band gap energies shown in Fig. 2. This means that absorption at these energies below the band gap of pure BiT yields mobile carriers that can participate in transport, as opposed to absorption that might come from fully localised defect states. Moreover, since there is increased absorption of visible light in BiT-LC owing to the reduced band gap, one can expect an enhanced total photoelectric response. Figure 4c shows $J(E)$ for BiT and 1B2L using a white light with a solar simulator. Both samples show a highly insulating behaviour in the dark. When the sample is illuminated, $J$ increases by approximately an order of magnitude even for pure BiT. A substantially greater enhancement (~35 fold increase of $J$) is observed from 1B2L. The enhancement of $J$ in BiT-LC is more obviously shown in Fig. 4d, where the time dependent $J$ shows a clear step-like increase and decrease by turning on and off the light, respectively. This clearly demonstrates the ability of absorbing more photons by the site-specific substitution, which can be in principle converted into electricity.

**Discussion**

Site-specific substitution in anisotropic oxides can offer an effective way to tuning band gaps in complex transition metal oxides. In particular, we showed the example of ferroelectric oxide $Bi_4Ti_3O_{12}$ where the band gap could be systematically controlled by site-specific substitutional alloying with $LaCoO_3$. This approach should not be limited to ferroelectrics but could be adapted to other oxide systems that have various exotic and useful characteristics. In addition, while this work has focused on site-specific substitution through the formation of a self-ordered superstructure, we envisage that our finding could be applied to other artificially-designed material systems, such as other Aurivillius phase materials, Ruddlesden-Popper series, and artificial oxide superlattices[29,30]. Thus, our approach may lead to opportunities for more successful integration of intriguing behaviours of complex oxides and other functionalities in electronic and energy applications.

## Methods

**Sample fabrication.** We have fabricated (001)-oriented BiT-LC epitaxial films (~40 nm in thickness) on single crystal (001) STO substrates using PLE with stoichiometric BiT and LCO targets. The thin films were grown with the heater temperature between 700 and 750 $^0$C in 100 mTorr of oxygen. A relatively high laser repetition rate of 10 Hz was used to prevent the loss of volatile Bi in the films. Quarter and half unit cell layers of LCO alternating with quarter unit cell layers of BiT were ablated for growing 1B1L and 1B2L alloys, respectively, by controlling the number of the laser pulses.

**STEM-EELS.** Cross sections of BiT and 1B2L epitaxial films were imaged in a fifth-order aberration corrected Nion UltraSTEM operated at 100 kV. Simultaneous high-angle ADF images and EELS were obtained to investigate the incorporation of La and Co in the BiT matrix. Elemental maps corresponding to the Ti-$L_{2,3}$, Co-$L_{2,3}$ and La-$M_{4,5}$ edges were produced by subtracting the background using a power law fit and integrating the remaining intensity under each edge in a 30 eV wide window. The acquisition time was 0.1s per pixel. EELS data are shown in Supplementary Fig. S3. Spectrum 1 is from the Bi plane in the $Bi_2O_2$ sub-layer of the compound showing enhanced segregation of La in the upper Bi planes. Spectrum 2 is from a Bi-O plane within the $Bi_2Ti_3O_{10}$ perovskite layers of the compound. Spectrum 3 is from a Ti-O plane near the $Bi_2O_2$ sub-layer showing a small signal of Co, at 779 eV (Co $L_3$ edge).

**Ferroelectric characterisation.** To evaluate the ferroelectric polarization, (104)-oriented epitaxial BiT and BiT-LC films were grown on (111) STO substrates covered with $SrRuO_3$ bottom electrodes using PLE. The polarization in such non-$c$-axis oriented thin films is aligned away from the film plane, so that the ferroelectric measurement can readily be performed in capacitor geometry[20]. Pt top electrodes (30 $\mu$m diameter) were deposited on the film surfaces using photolithographic lift-off to fabricate capacitors. The PE hysteresis loops were recorded with a standard TF analyzer at 100-500 Hz.

**Ellipsometry.** The optical properties of the (001) BiT-LC samples were investigated using a two-modulator generalised ellipsometer[31,32] in the wavelength range between 232 and 850 nm (1.46 − 5.34 eV) at four angles of incidence (55, 60, 65, and 70°). The near surface region was modelled using a four-medium model: air/surface roughness/film (BiT-LC)/substrate (STO). The surface roughness (typically less than 5 nm) was modelled using the Bruggeman effective medium approximation[33] consisting of 50% film and 50% voids. (An interface layer was included in some of the calculations, but was not found to improve the fits). The ellipsometric spectra taken at multiple angles of incidence were then used to determine the spectroscopic dielectric functions $\varepsilon_1(\omega)$ and $\varepsilon_2(\omega)$ of the films, as well as the film thickness. The film thickness was ~40 nm, which was consistent with the values obtained from low angle X-ray reflection measurements. The data is reported as the optical conductivity $\sigma_1 = \varepsilon_2\omega/4\pi$.

**Photoconductance measurements.** Interdigitated Pt electrodes with a 10 $\mu$m separation were patterned using the photolithography lift-off process on (001) BiT-LC films for in-plane photoconductance measurements. The electrodes were then wired using gold wires. Light was illuminated using either a solar simulator (250 W) or a monochromator. For the monochromator, the number of incident photons was determined using a NIST-calibrated photodiode. The current density is then normalised to the number of photons for each wavelength.

**Density functional calculations.** We constructed a supercell based on the two formula unit primitive cell of $Bi_4Ti_3O_{12}$ substituting one Ti by Co and removing one O as described in the text to yield a formula $Bi_8Ti_5CoO_{23}$. The internal atomic positions were then fully relaxed including the spin polarization, so that Co could carry a moment. The calculation was done within density functional theory using the augmented planewave plus local orbital extension of the general potential linearised augmented planewave (LAPW) method as implemented in the wien2k code. The initial relaxation was done using the generalised gradient approximation of Perdew, Burke and Ernzerhof (PBE) with a scalar relativistic approximation. We tested our calculations using the PBE+$U$ method with various values of the Coulomb parameter $U$; the results were insensitive to $U$ provided that it is large enough to remove the Co $d$ states from the band gap. The reason is that the band edge states induced by the substitution come mainly from Bi.

# References

1. Capasso, F. Band-Gap Engineering: From Physics and Materials to New Semiconductor Devices. *Science* **235**, 172-176 (1987).
2. Ramirez, A.P. Oxide Electronics Emerge. *Science* **315**, 1377-1378 (2007).
3. Takagi, H. & Hwang, H.Y. An Emergent Change of Phase for Electronics. *Science* **327**, 1601-1602 (2010).
4. Kozuka, Y., *et al.* Two-dimensional normal-state quantum oscillations in a superconducting heterostructure. *Nature* **462**, 487-490 (2009).
5. Tsukazaki, A., *et al.* Observation of the fractional quantum Hall effect in an oxide. *Nature Mater.* **9**, 889-893 (2010).
6. Lewis, B.G. & Paine, D.C. Applications and processing of transparent conducting oxides. *MRS Bull.* **25**, 22-27 (2000).
7. Choi, T., Lee, S., Choi, Y.J., Kiryukhin, V. & Cheong, S.-W. Switchable Ferroelectric Diode and Photovoltaic Effect in $BiFeO_3$. *Science* **324**, 63-66 (2009).
8. Yang, S.Y., *et al.* Above-bandgap voltages from ferroelectric photovoltaic devices. *Nature Nanotechnol.* **5**, 143-147 (2010).
9. Yuan, Y., *et al.* Efficiency enhancement in organic solar cells with ferroelectric polymers. *Nature Mater.* **10**, 296-302 (2011).
10. Bennett, J.W., Grinberg, I. & Rappe, A.M. New Highly Polar Semiconductor Ferroelectrics through $d^8$ Cation-O Vacancy Substitution into $PbTiO_3$: A Theoretical Study. *J. Am. Chem. Soc.* **130**, 17409-17412 (2008).
11. Huang, H. Solar energy: Ferroelectric photovoltaics. *Nature Photon.* **4**, 134-135 (2010).
12. Ehara, S., *et al.* Dielectric Properties of $Bi_4Ti_3O_{12}$ below the Curie Temperature. *Jpn. J. Appl. Phys.* **20**, 877-881 (1981).
13. Singh, D.J., Seo, S.S.A. & Lee, H.N. Optical properties of ferroelectric $Bi_4Ti_3O_{12}$. *Phys. Rev. B* **82**, 180103 (2010).
14. Jia, C., Chen, Y. & Zhang, W.F. Optical properties of aluminum-, gallium-, and indium-doped $Bi_4Ti_3O_{12}$ thin films. *J. Appl. Phys.* **105**, 113108-113104 (2009).

15. Rae, A.D., Thompson, J.G., Withers, R.L. & Willis, A.C. Structure refinement of commensurately modulated bismuth titanate, $Bi_4Ti_3O_{12}$. *Acta Cryst.* **B46**, 474-487 (1990).

16. Cummins, S.E. & Cross, L.E. Crystal symmetry, optical properties, and ferroelectric polarization of $Bi_4Ti_3O_{12}$ single crystals. *Appl. Phys. Lett.* **10**, 14-16 (1967).

17. Scott, J.F. *Ferroelectric Memories*, (Springer, Heidelberg, Germany, 2000).

18. Shimakawa, Y., *et al.* Crystal and electronic structures of $Bi_{4-x}La_xTi_3O_{12}$ ferroelectric materials. *Appl. Phys. Lett.* **79**, 2791-2793 (2001).

19. Park, B.H., *et al.* Lanthanum-substituted bismuth titanate for use in non-volatile memories. *Nature* **401**, 682-684 (1999).

20. Lee, H.N., Hesse, D., Zakharov, N. & Gösele, U. Ferroelectric $Bi_{3.25}La_{0.75}Ti_3O_{12}$ Films of Uniform *a*-Axis Orientation on Silicon Substrates. *Science* **296**, 2006-2009 (2002).

21. Arima, T., Tokura, Y. & Torrance, J.B. Variation of optical gaps in perovskite-type 3*d* transition-metal oxides. *Phys. Rev. B* **48**, 17006 (1993).

22. Hashimoto, T. & Moriwake, H. Oxygen vacancy formation energy and its effect on spontaneous polarization in $Bi_4Ti_3O_{12}$ : A first-principles theoretical study. *Phys. Rev. B* **78**, 092106 (2008).

23. Noguchi, Y., Soga, M., Takahashi, M. & Miyayama, M. Oxygen stability and leakage current mechanism in ferroelectric La-substituted $Bi_4Ti_3O_{12}$ single crystals. *Jpn. J. Appl. Phys.* **44**, 6998-7002 (2005).

24. Lee, H.N. & Hesse, D. Anisotropic ferroelectric properties of epitaxially twinned $Bi_{3.25}La_{0.75}Ti_3O_{12}$ thin films grown with three different orientations. *Appl. Phys. Lett.* **80**, 1040-1042 (2002).

25. Seo, S.S.A., *et al.* Multiple conducting carriers generated in $LaAlO_3/SrTiO_3$ heterostructures. *Appl. Phys. Lett.* **95**, 082107 (2009).

26. Lee, H., *et al.* Dielectric functions and electronic band structure of lead zirconate titanate thin films. *J. Appl. Phys.* **98**, 094108 (2005).

27. Moret, M.P., Devillers, M.A.C., Worhoff, K. & Larsen, P.K. Optical properties of $PbTiO_3$, $PbZr_xTi_{1-x}O_3$, and $PbZrO_3$ films deposited by metalorganic chemical vapor on $SrTiO_3$. *J. Appl. Phys.* **92**, 468-474 (2002).

28. Hu, Z.G., Li, Y.W., Yue, F.Y., Zhu, Z.Q. & Chu, J.H. Temperature dependence of optical band gap in ferroelectric $Bi_{3.25}La_{0.75}Ti_3O_{12}$ films determined by ultraviolet transmittance measurements. *Appl. Phys. Lett.* **91**, 221903 (2007).

29. Haeni, J.H., *et al.* Epitaxial growth of the first five members of the Sr$_{n+1}$Ti$_n$O$_{3n+1}$ Ruddlesden--Popper homologous series. *Appl. Phys. Lett.* **78**, 3292-3294 (2001).

30. Lee, H.N., Christen, H.M., Chisholm, M.F., Rouleau, C.M. & Lowndes, D.H. Strong polarization enhancement in asymmetric three-component ferroelectric superlattices. *Nature* **433**, 395-399 (2005).

31. Jellison, G.E. & Modine, F.A. Two-modulator generalized ellipsometry: experiment and calibration. *Appl. Opt.* **36**, 8184-8189 (1997).

32. Jellison, G.E. & Modine, F.A. Two-modulator generalized ellipsometry: theory. *Appl. Opt.* **36**, 8190-8198 (1997).

33. Bruggeman, D.A.G. Berechnung verschiedener physikalischer Konstanten von heterogenen Substanzen. I. Dielektrizitätskonstanten und Leitfähigkeiten der Mischkörper aus isotropen Substanzen. *Annalen der Physik* **416**, 636-664 (1935).

**Acknowledgements**

We thank J. F. Scott and S. S. A. Seo for discussions. This work was supported by the U.S. Department of Energy, Basic Energy Sciences, Materials Sciences and Engineering Division (synthesis and electrical characterization) and the Laboratory Directed Research and Development Program of Oak Ridge National Laboratory (STEM-EELS, theory and optical characterization). The optical measurement was in part conducted at the Center for Nanophase Materials Sciences, a DOE-BES user facility.

**Author contributions**

Sample synthesis by PLE and structural and electrical characterisation were carried out by W.S.C. under the guidance of H.N.L. Spectroscopic ellipsometry and photocurrent experiments were performed by W.S.C., T. C. and G.E.J. Z-contrast STEM and EELS experiments were conducted by M.F.C. The theoretical calculations were performed by D.J.S. The project was planned and managed by H.N.L. All authors contributed to the manuscript writing.

## Additional information

**Supplementary Information** accompanies this paper

**Competing financial interests:** The authors declare no competing financial interests.

## Figure Legends

**Figure 1 | Effects of site-specific substitution on structural and ferroelectric properties.** (**a**) Schematic diagram of a half pseudo-orthorhombic unit cell of BiT. The arrows indicate the most probable sites where BiT is substituted by LCO. (**b**) X-ray $\theta$-$2\theta$ diffraction patterns of BiT (blue), 1B1L (green), and 1B2L (red) thin films. The 001 peak from STO substrates and the separation of the 004 peak in BiT-LC are indicated as * and arrows, respectively. (**c**) X-ray rocking curve of the 008 reflection of 1B2L (FWHM = 0.025°). (**d**) Atomic resolution Z-contrast STEM images are shown for BiT and 1B2L. The overall layered structure of BiT is well preserved despite the incorporation of LC, indicating that some portions of BiT are substituted by La and Co without deteriorating the layered structure. The scale bars correspond to 1 nm. (**e**) Annular dark field image (ADF) and corresponding elemental maps for Ti and La visualised by EELS from a 1B2L sample. La elemental map shows that La mostly substitutes for Bi in the upper layer of the $Bi_2O_2$ layer. (**f**) $P(E)$ hysteresis loops of BiT (blue) and BiT-LC (red).

**Figure 2 | Optical property and band gap change.** (**a**) Optical conductivity ($\sigma_1(\omega)$) of BiT (blue), 1B1L (green), 1B2L (red), and LCO (grey) show a systematic and substantial decrease in band gap upon site-specific substitution of BiT with La and Co. The thickness of the spectra corresponds to the fitting error bar. Note that the non-zero $\sigma_1(\omega)$ below the band gap (particularly for 1B2L) may be an artefact, originating from ellipsometry fitting procedure. The extrapolation lines (dashed lines) indicate optical band gaps ($E_g$) of ~3.55, ~3.30, and ~2.65 eV for BiT, 1B1L, and 1B2L, respectively. The decreased band gap is mainly due to a state formed below the original band gap of BiT, as shown by the red arrow. The optical band gap of 1B2L is green, well within the solar spectrum. (**b**) Absorption coefficient ($\alpha(\omega)$) of typical semiconductors [GaAs (orange), CdTe (green), crystalline Si (blue), AlAs (purple)] (data taken from the dielectric constant database of WVASE32 (J. A. Woollam Co.)). $\alpha(\omega)$ of STO (black) and LCO (thick grey) (measured using ellipsometry) are also shown. LCO has the advantage of a lower band gap and a higher $\alpha(\omega)$ as compared to the typical semiconductors. Photographs of (**c**) BiT and (**d**) 1B2L samples intuitively showing the effect of alloying on the band gap.

**Figure 3 | Electronic structure.** (**a**) Calculated total electronic density of states (DOS) (blue) and projection onto the Bi (green) and Ti (orange) atoms for BiT with PBE GGA. The obtained band gap is 2.34 eV (~1.2 eV smaller than the experimental result since standard density functional theory underestimates the gap value.) (**b**) Calculated total electronic DOS (blue) and projection onto the Bi (green) and Co (red) atoms for the Co:BiT ($Bi_8Ti_5CoO_{23}$ supercell) with the PBE GGA and the PBE+$U$ method with $U$-$J$=5 eV for Co atom. Majority and minority spins are shown as positive and negative, respectively. The obtained band gap for the Co:BiT is 1.64 eV, resulting in ~30% decrease in the band gap. The theoretical result is in accord with the experimental observation. The origin of the reduced band gap is additional states below the conduction band of BiT, indicated as a red arrow. This enables charge transfer excitations below the band gap of BiT, shown as the same red arrow in Fig. 2.

**Figure 4| Optoelectric properties.** (**a**) Photon energy dependence of $J(E)$ of a 1B2L film measured by illuminating light of various wavelength using a monochromator. $J$ is normalised to the number of photons at each wavelength. (**b**) Summary of the photoconductance as a function of applied photon energy for BiT (blue) and 1B2L (red) at 10 kV/cm. Horizontal dashed lines indicate the dark current for each sample. The inset shows magnified version of the dotted squared region. Arrows indicate the onsets of photocunductance at 3.10 and 2.75 eV for BiT and 1B2L, respectively. (**c**) $J(E)$ curves from BiT (blue) and 1B2L (red) thin films. Illumination of white light by a solar simulator (250 W) enhanced the current level by ~10 and ~35 times for BiT and 1B2L, respectively, indicated by arrows. (**d**) Photocurrent density recorded from BiT (blue) and 1B2L (red) thin films by switching on and off the solar simulator at 20 kV/cm.

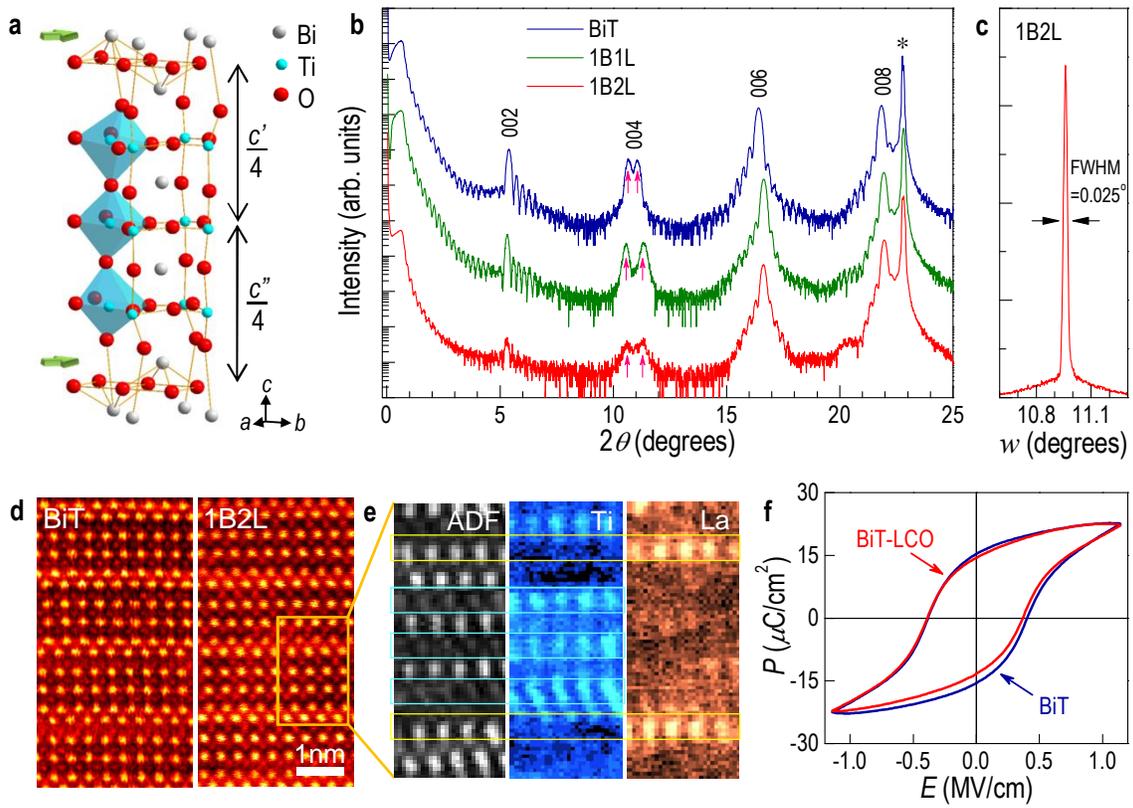

Fig. 1
W. S. Choi *et al.*

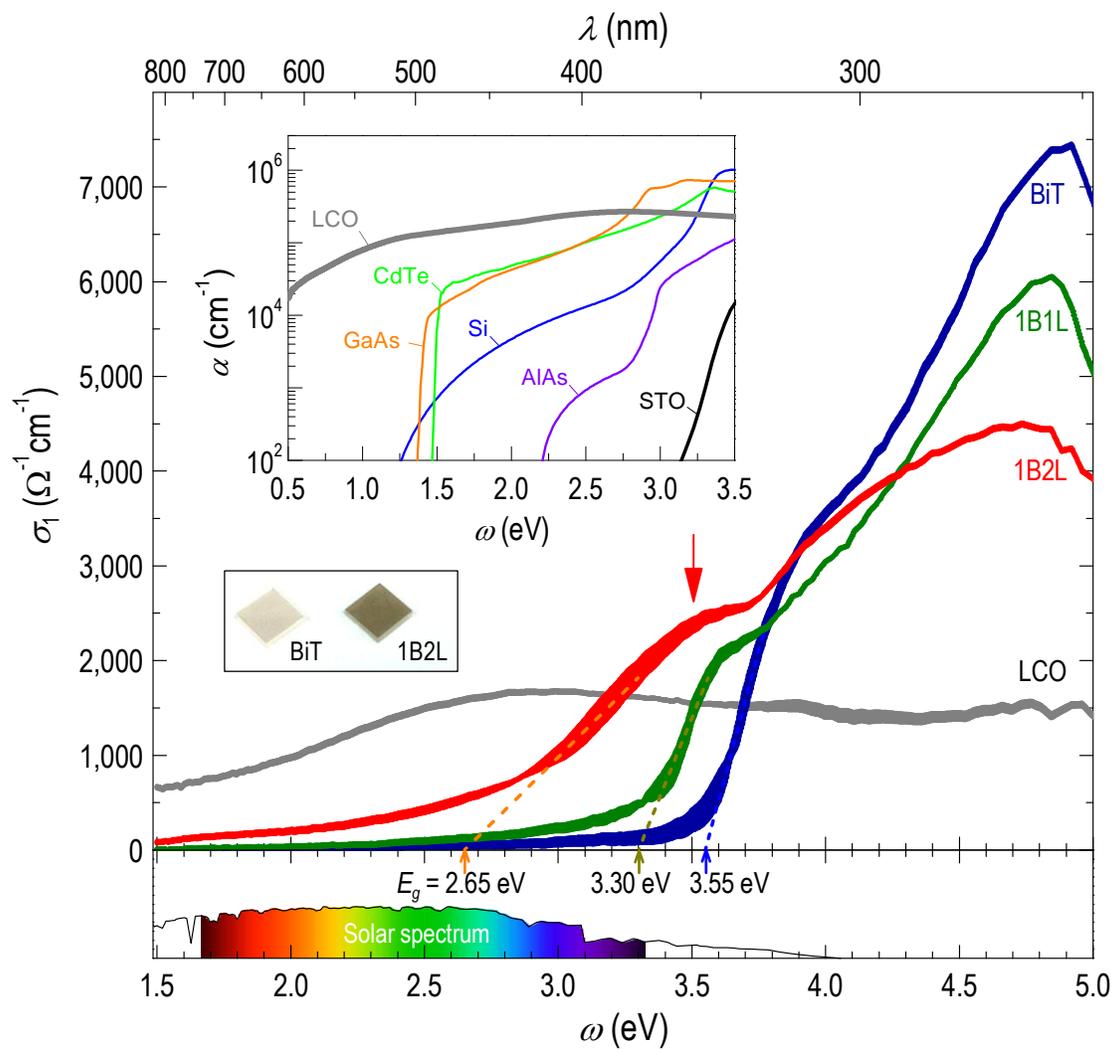

Fig. 2
W. S. Choi *et al*.

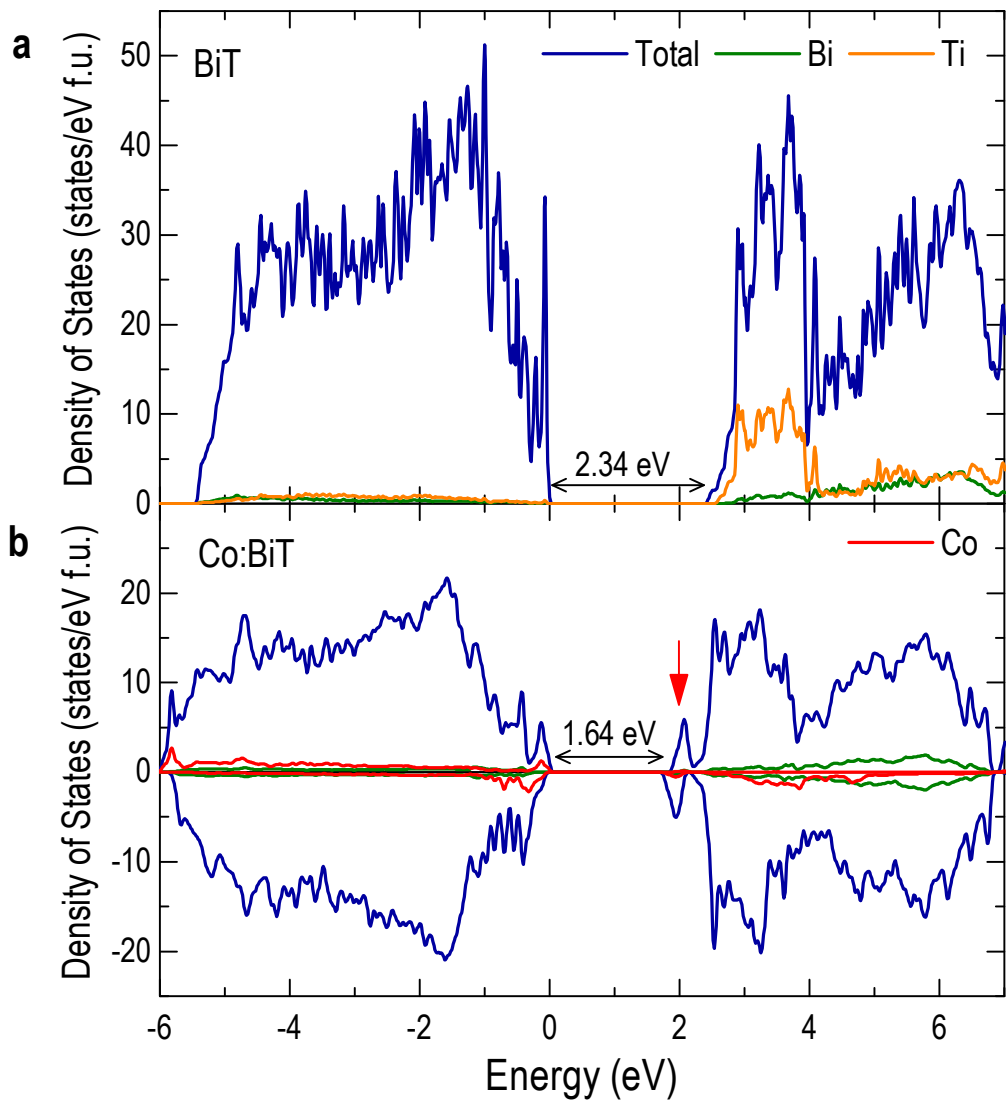

Fig. 3
W. S. Choi *et al.*

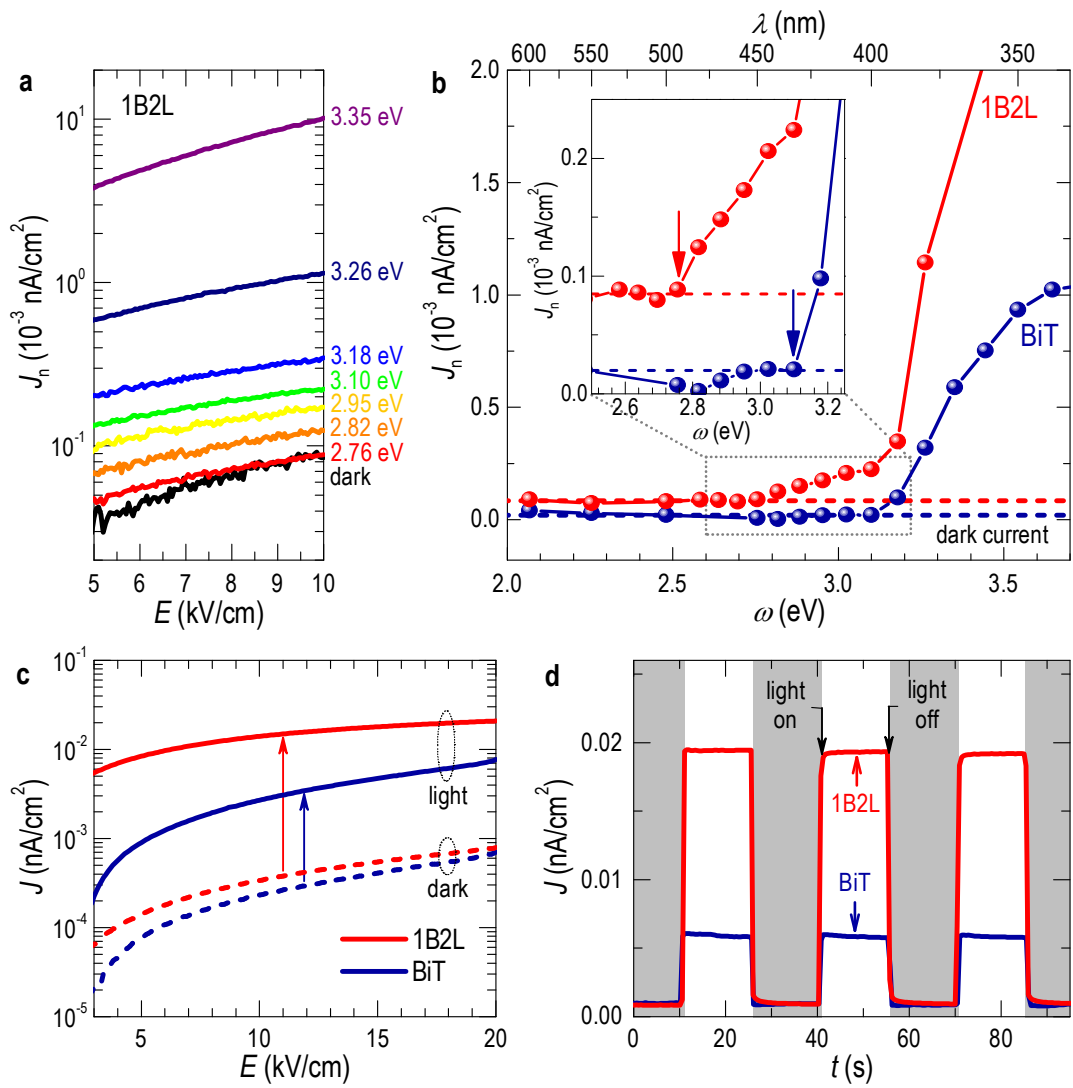

Fig. 4
W. S. Choi *et al.*

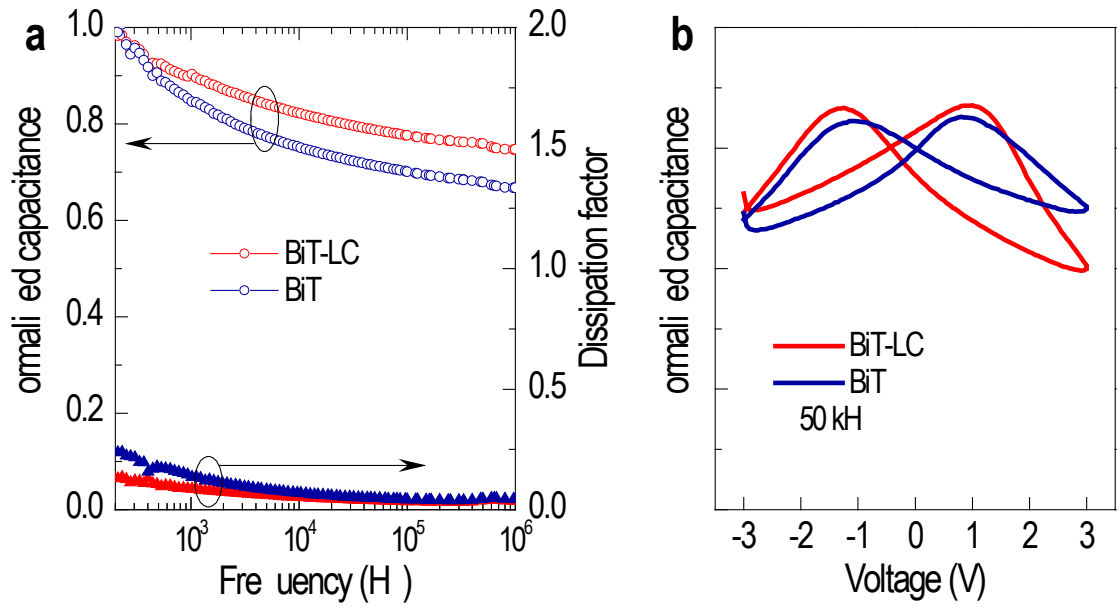

**Supplementary Figure S1 | Ferroelectric and dielectric properties.** (**a**) *C-f* and tan$\delta$ and (**b**) *C-V* curves of (104)-oriented BiT and 1B2L films. The capacitance is normalized due to the different film thickness. The calculated dielectric constants are around 85.

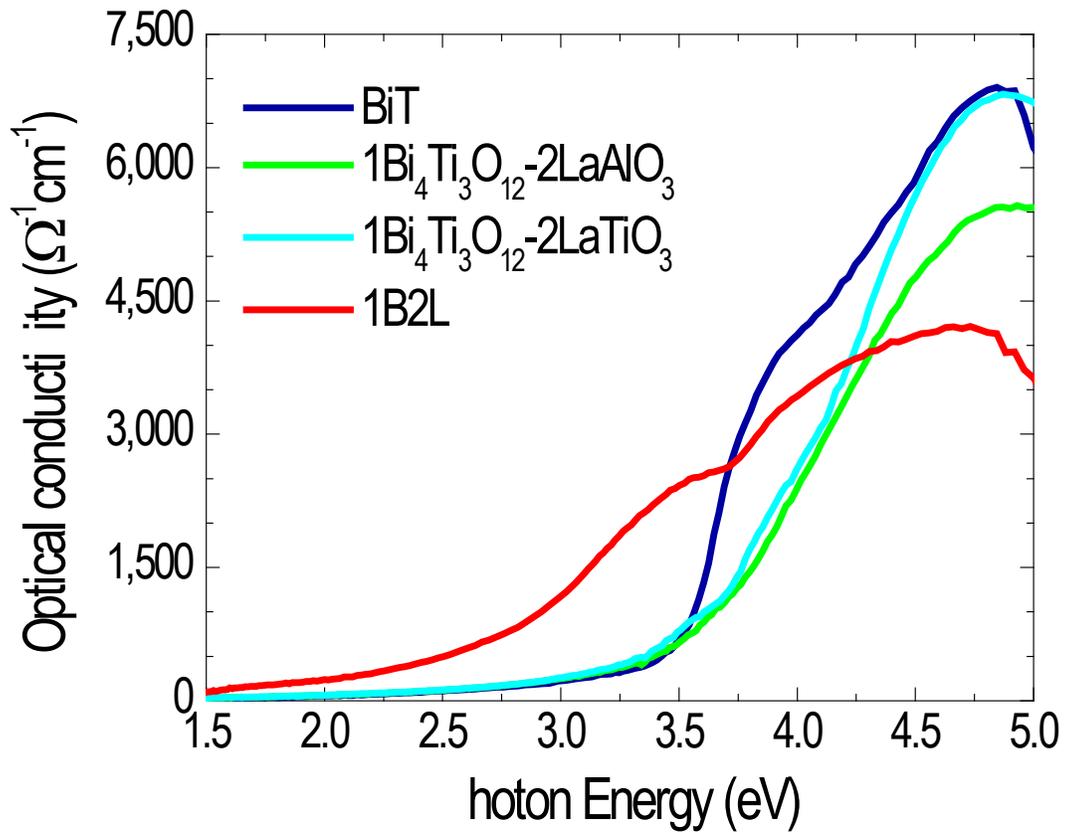

**Supplementary Figure S2 | Optical conductivity spectra of $Bi_4Ti_3O_{12}$ doped with different La*TM*O$_3$.** $LaAlO_3$ and $LaTiO_3$ are used as alloying materials instead of LCO for 1BiT-2LaAlO$_3$ and 1BiT-2LaTiO$_3$ films. Both films have similar band gaps as pure BiT, showing the importance of Co in decreasing the band gap in BiT.

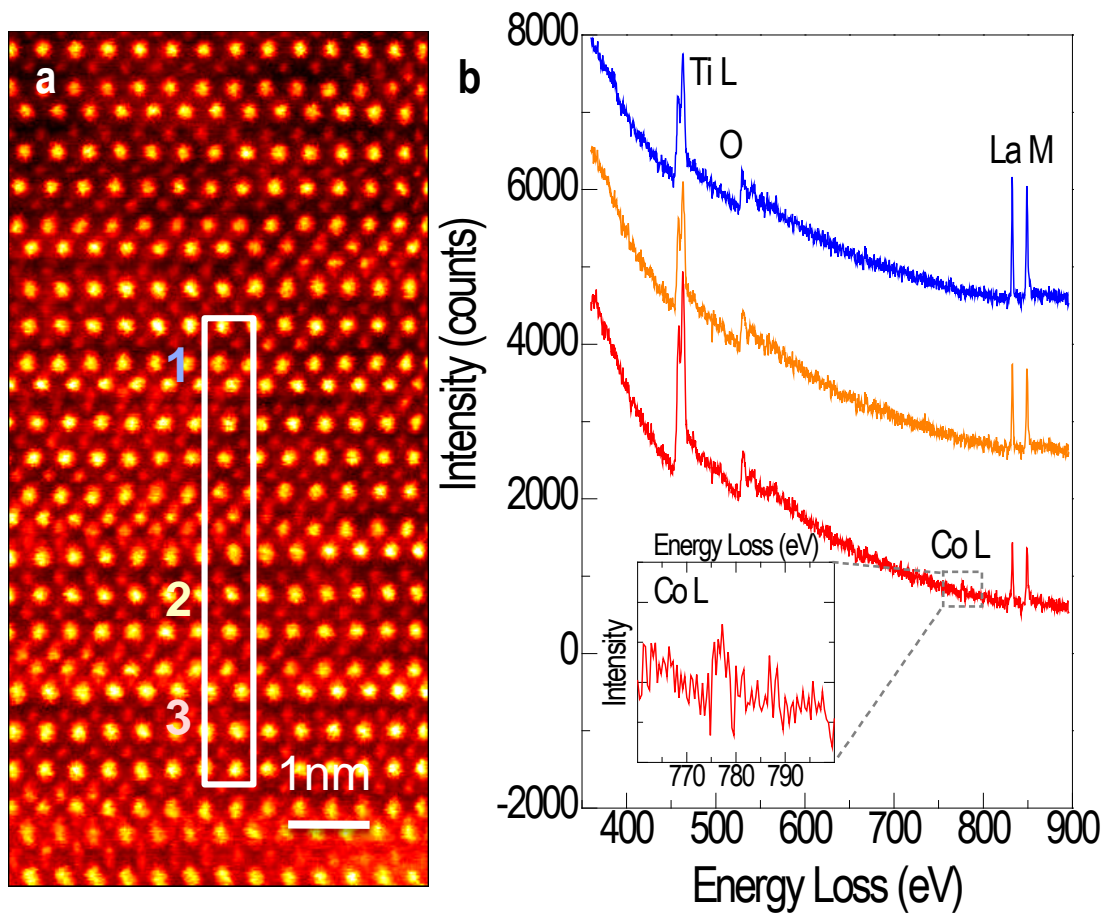

**Supplementary Figure S3 | Atomic-scale imaging and spectroscopy.** (**a**) Cross-sectional Z-contrast STEM image of 1B2L. The growth direction of the imaged film is from the bottom to the top of the image. (**b**) EELS data corresponding to the numbered spots in the left panel. The *y*-axis is in arbitrary units with spectrum 2 and 1 offset upwards by 2000 and 4000 counts, respectively. Spectrum 1 is from the Bi plane in the $Bi_2O_2$ sub-layer of the compound showing enhanced segregation of La in the upper Bi planes. Spectrum 2 is from a Bi-O plane within the $Bi_2Ti_3O_{10}$ perovskite layers of the compound. Spectrum 3 is from a Ti-O plane near the $Bi_2O_2$ sub-layer showing a small signal of Co, at 779 eV (Co $L_3$ edge). The inset shows magnified Co L-edge showing small signal of Co.

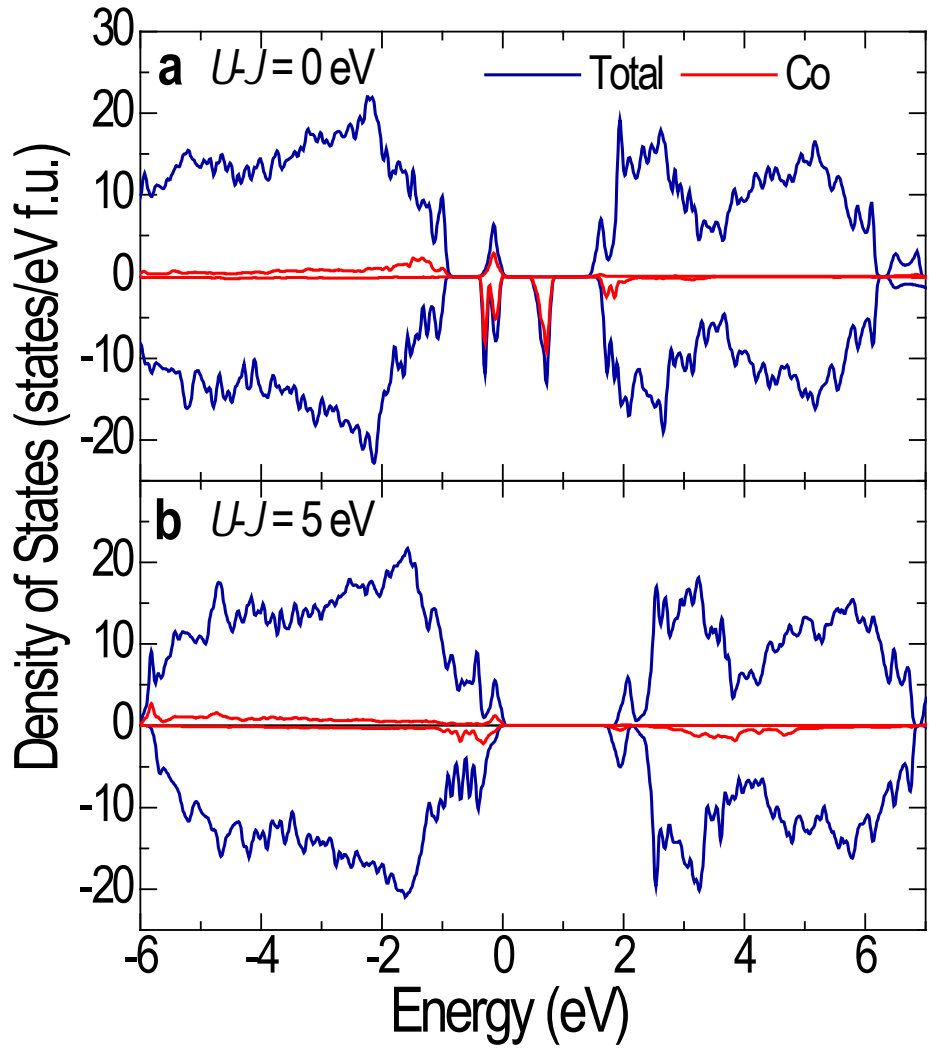

**Supplementary Figure S4 | Density functional theory calculation.** Electronic density of states of Co:BiT by varying $U - J$ value. (**a**) $U - J = 0$ eV and (**b**) $U - J = 5$ eV (the same as Fig. 3b)

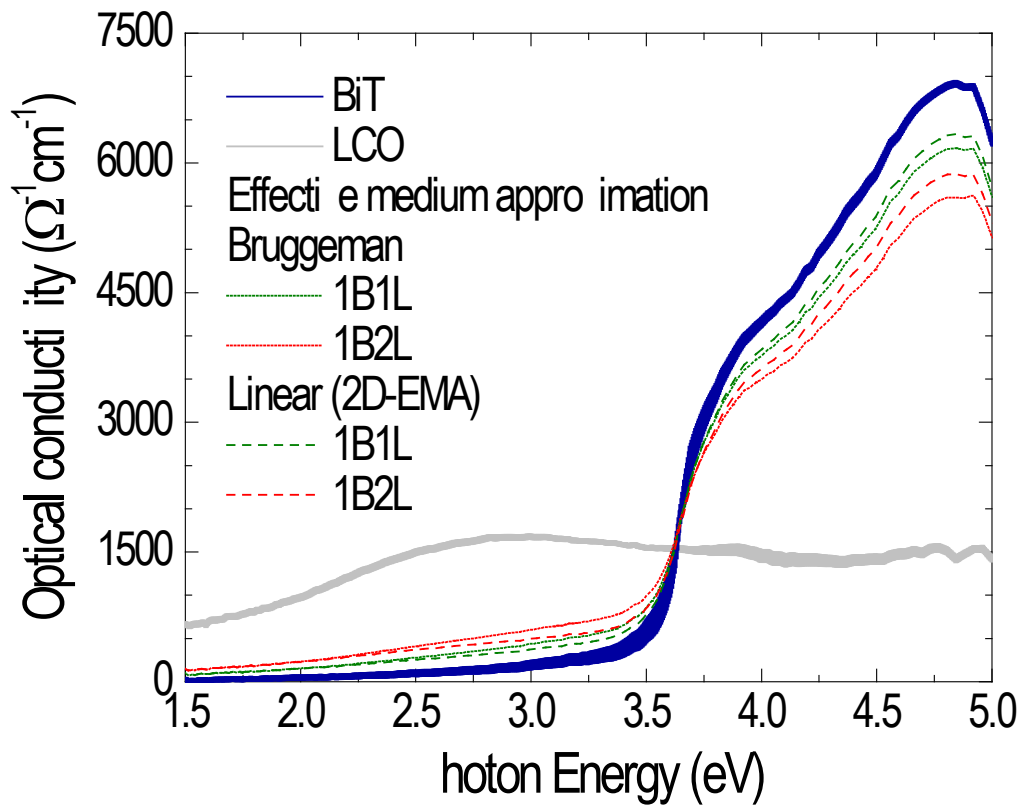

**Supplementary Figure S5 | Effective medium approximation results.** The EMA spectra are constructed based on the optical conductivity spectra of the constituent material: BiT and LCO. The result does not reproduce the observed experimental spectra shown in Fig. 2, validating the substantial modification of the electronic structure in our BiT-LC thin films.

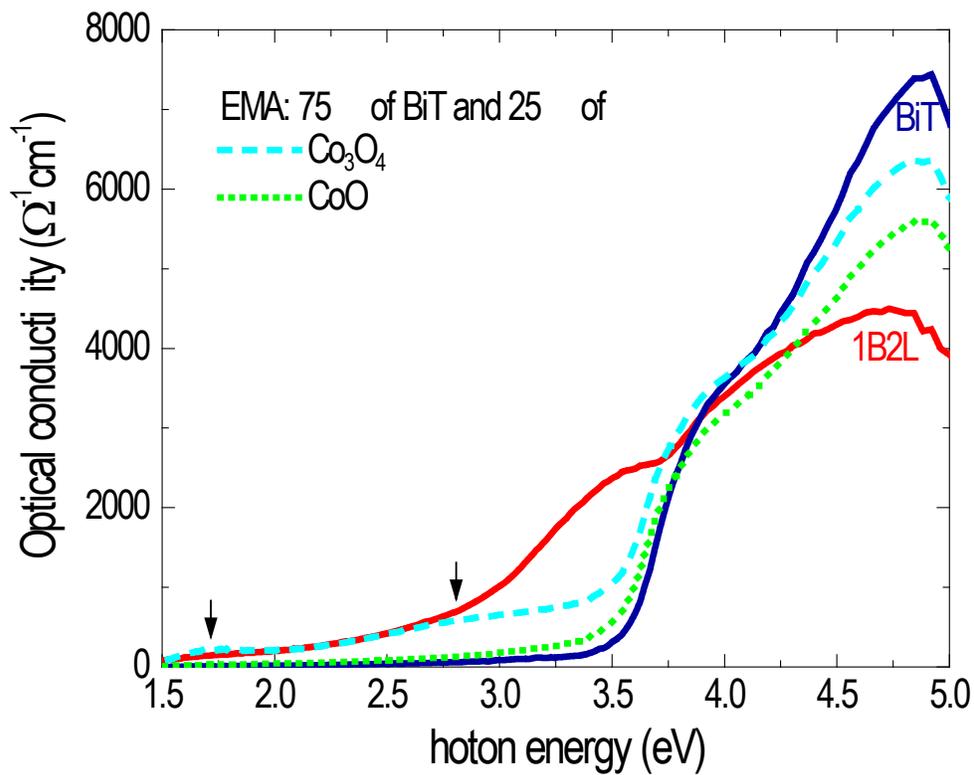

**Supplementary Figure S6 | Simulated optical conductivity spectra with cobalt oxide impurities.** The spectra were obtained from the Bruggeman EMA for mixtures of BiT and cobalt oxides. The volume fraction of the cobalt oxides was set to 25% of the whole composite. The weak peak features at ~1.7 and ~2.8 eV in EMA of BiT-$Co_3O_4$ are indicated as black arrows. The spectra of BiT and 1B2L are also shown as comparison.

**Ferroelectric and dielectric measurements.** Ferroelectric properties including capacitance and dissipation factor are measured. As shown in Supplementary Fig. S1a, the capacitance and dissipation factor of BiT are slowly decreasing with increasing frequency and the dissipation factor is around 0.03 at high frequencies, consistent with the reported values[34-36]. BiT-LC shows a similar behavior and value. The voltage dependent capacitance (Supplementary Fig. S1b) show well-defined butterfly-like curves, confirming good ferroelectricity in both BiT and BiT-LC.

**Theoretical Calculation.** Discussing the density functional theory (DFT) in additional detail, we chose $Co^{2+}$ instead of $Co^{3+}$ since more stable divalent Co compounds have band gaps in the range observed in the ellipsometry result as compared to $Co^{3+}$ compounds, such as $LaCoO_3$. Furthermore, while $Co^{3+}$ would require a rather complex defect structure, $Co^{2+}$ requires simply one O vacancy along with the substitution of $Ti^{4+}$ with $Co^{2+}$ to satisfy the charge balance. It should also be noted that $Co^{4+}$ oxides such as de-intercalated $Li_xCoO_2$ or $SrCoO_3$ are chemically unstable and reactive. Therefore, a supercell of $Bi_8Ti_5CoO_{23}$ was constructed and then fully relaxed the internal atomic positions. As mentioned before, the relaxation was done including spin polarization so that Co could carry a moment. We indeed found out that this yields $Co^{2+}$ with a very stable magnetic moment of 3 $\mu_B$ per Co and a substantial gap in the Co $d$ spectrum even with the standard PBE functional ($U = 0$). The spin polarization of the host is due to the fact that there is only one Co atom in the supercell. This weak spin polarization would be averaged out in the paramagnetic alloy. Note that the Co is five-fold coordinated which yields a different crystal field scheme than an octahedral coordination. The four in-plane Co-O bond lengths range from 1.92 Å to 2.04 Å, while the single apical O (towards the center of the perovskite block) distance is 2.36 Å. The Co bond-valence sum for this site is 1.88, which is a reasonable value for $Co^{2+}$ in a chemically stable compound. The electronic density of states is shown in Supplementary Fig. S4a for this calculation of PBE. With $U-J=0$ we find a gap of 0.48 eV of $d-d$ character at the PBE level. The $d$-states in the gap are no doubt an artifact due to the neglect of the Coulomb repulsion on Co. As may be seen, there is a strong spin dependent hybridization of Co states in the O $2p$ derived valence bands. Importantly, there is a split-off state from the cation derived conduction bands. We also did PBE+$U$ calculations with different values of $U-J$. Supplementary Figure S4b (the same as in Fig. 3b) shows the PBE+$U$ calculation result with $U-J=5$ eV applied to Co only. For values appropriate to 3$d$ transition metal oxides as shown in Supplementary Fig. S4b, we find that the mid-gap *d*-states are in fact removed and become strongly mixed into the valence and conduction bands. The shape of the density of states is modified. However, in common with the *U-J*=0 calculation, we find a split-off density of states below the conduction band edge. This finding is insensitive to the value of *U-J*. This split-off density of states means that there will be charge transfer excitations below the pure compound gap consistent with our optical data.

**Effective Medium Approximation (EMA).** In general, when an external material is introduced to a host material preserving their optical constants, the optical property of the whole material can be obtained from an effective medium approximation (EMA). We used two different EMA methods (Bruggeman[37] and linear[38]) to test this idea as shown in Supplementary Fig. S5. Based on the pure BiT and LCO dielectric spectra, we constructed optical conductivity spectra for 1B1L and 1B2L. Note that the nominal volume ratios of LCO within the samples are 11.9% and 23.7%, respectively, for 1B1L and 1B2L. As shown in Supplementary Fig. S5, neither EMA methods could reproduce our experimental optical conductivity spectra. In particular, we did not observe any additional states that cause the band gap reduction. The result, therefore, indicates the substantial modification of the electronic structure as shown in theoretical calculation is necessary to satisfy the observed band gap change.

**Elimination of unexpected Co phases as origin of the reduced band gap.** It has been known that Co easily forms cobalt oxide phases (Usually $Co_3O_4$ or CoO). We carefully examined the possible formation of such Co-based secondary phases and their influence to our optical conductivity spectra. First of all, from XRD measurements, we could not see any of the secondary phases. Second, if such phases were to be formed, the band gap would not decrease systematically with increasing LCO incorporation. Rather, the absorption will increase above the modified band gap. To further confirm that the cobalt oxide phases are not contributing to the reduced band gap, we used the following EMA analyses. We tested Bruggeman EMA[S1] using optical properties of BiT, $Co_3O_4$, and CoO[39-41], as shown in Supplementary Fig. S6. Linear EMA gave very similar results. The results clearly showed that the band gap reduction could not be realized, even if there exist a significant amount of such cobalt oxides within the BiT film.

Specifically, the peak at 3.4 eV in 1B2L, and the band gap change due to this absorption could not be observed in the EMA. Also note that the EMA of BiT and $Co_3O_4$ showed some absorption at low photon energy (shown as arrows). This is due to the low-lying absorption of $Co_3O_4$ at ~1.7 and ~2.8 eV. The absence of such peaks in 1B2L affirms that the contribution of $Co_3O_4$ is very minuscule (if there is any) to the optical property of our 1B2L film, unambiguously validating the effect of Co substitution in BiT on the band gap reduction.

34. Joshi, P. C. & Desu, S. B. Structural and electrical characteristics of rapid thermally processed ferroelectric $Bi_4Ti_3O_{12}$ thin films prepared by metalorganic solution deposition technique. *J. Appl. Phys.* **80**, 2349–2357 (1996).
35. Kim, J. S., Kim, S. S., Kim, J. K. & Song, T. K. Ferroelectric properties of tungsten-substituted $Bi_4Ti_3O_{12}$ thin film prepared by sol-gel method. *Jpn. J. Appl. Phys.* **41**, 6451–6154 (2002).
36. Hou, R. Z. & Chen, X. M. Dielectric properties of $Bi_{4-x}La_xTi_3O_{12}$ ($0 \leq x \leq 2$) ceramics. *J. Electroceram.* **10**, 203–207 (2003).
37. Bruggeman, D. A. G. Berechnung verschiedener physikalischer Konstanten von heterogenen Substanzen. *Ann. Phys*. **24**, 636–679 (1935).
38. Choi, W. S., Ohta, H., Moon, S. J., Lee, Y. S. & Noh, T. W. Dimensional crossover of polaron dynamics in $Nb:SrTiO_3/SrTiO_3$ superlattices: Possible mechanism of thermopower enhancement. *Phys. Rev. B* **82**, 024301 (2010).
39. Powell, R. J. & Spicer, W. E. Optical properties of NiO and CoO. *Phys. Rev. B* **2**, 2182–2193 (1970).
40. Cook, J. G. & Van der Meer, M. P. The optical properties of sputtered $Co_3O_4$ films. *Thin Solid Films* **144**, 165–176 (1986).
41. Ruzakowski Athey, P., Urban III, F. K., Tabet, M. F. & McGahan, W. A. Optical properties of cobalt oxide films deposited by spray pyrolysis. *J. Vac. Sci. Technol. A* **14**, 685–692 (1996).